# Observation of superconductivity and enhanced upper critical field of $\eta$-carbide-type oxide Zr$_4$Pd$_2$O

Yuto Watanabe[1,*], Akira Miura[2], Chikako Moriyoshi[3], Aichi Yamashita[1] and Yoshikazu Mizuguchi[1,*]

[1]Department of Physics, Tokyo Metropolitan University, Hachioji, Tokyo192-0397, Japan.
[2]Faculty of Engineering, Hokkaido University, Sapporo, Hokkaido 060-8628, Japan.
[3]Graduate School of Advanced Science and Engineering, Hiroshima University, Higashihiroshima, Hiroshima 739-8526, Japan
*watanabe-yuto@ed.tmu.ac.jp
*mizugu@tmu.ac.jp

## ABSTRACT

We report the first observation of bulk superconductivity of a $\eta$-carbide-type oxide Zr$_4$Pd$_2$O. The crystal structure and the superconducting properties were studied through synchrotron X-ray diffraction, magnetization, electrical resistivity, and specific heat measurement. The superconducting transition was observed at $T_c$ = 2.73 K. Our measurement revealed that the $\eta$-carbide-type oxide superconductor Zr$_4$Pd$_2$O shows an enhanced upper critical field $\mu_0 H_{c2}(0)$ = 6.72 T, which violates the Pauli-Clogston limit $\mu_0 H_P$ = 5.29 T. On the other hand, we found that the enhanced upper critical field is absent in a Rh analogue Zr$_4$Rh$_2$O. The large $\mu_0 H_{c2}(0)$ of Zr$_4$Pd$_2$O would be raised from strong spin-orbit coupling with Pd-4$d$ electrons. The discovery of new superconducting properties for Zr$_4$Pd$_2$O would shed light on the further development of $\eta$-carbide-type oxide superconductors.

## Introduction

Transition metal oxides are well known as one of the most fascinating solids because of their variety of physical properties [1] such as metal-insulator transition [2], giant magnetoresistance [3], ferroelectricity [4], and high-temperature superconductivity [5]. In transition metal oxides, anisotropic-shaped $d$-orbital electrons and strong electron correlation effect caused by Coulomb interaction between electrons play important roles for emerging the various physical phenomena [6]. The first discovery of superconductors in oxide compounds is a perovskite-type structural SrTiO$_{3-x}$ [7], and the discovery has led to the development of many kinds of oxide superconductors, for example, Ba$_{1-x}$K$_x$BiO$_3$ [8], YBa$_2$Cu$_3$O$_7$ [9], Li$_{1+x}$Ti$_{2-x}$O$_4$ [10] and so on. Among the discoveries of oxide superconductors, the $\eta$-carbide-type superconductors A$_4$B$_2$X have attracted attention in recent studies; here, A and B are transition metals and X is a light element such as carbon, nitrogen, or oxygen [11,12]. Ma et al. reported the bulk superconductivity of Zr$_4$Rh$_2$O$_x$ ($x$ = 0.7 and 1) [12] and Nb$_4$Rh$_2$C$_{1-\delta}$ [13] at transition temperatures of $T_c$ = 2.8 K ($x$ = 0.7), 4.8 K ($x$ = 1) and 9.75 K, respectively. They also found that Ti$_4$M$_2$O with M = Co, Rh, and Ir show superconductivity at $T_c$ = 2.7 K, 2.8 K, and 5.4 K, respectively [14]. The significant discovery of them was not only founding superconductors but also observing a large upper critical field $\mu_0 H_{c2}(0)$ violating the Pauli-Clogston limit (Pauli limit) for Nb$_4$Rh$_2$C$_{1-\delta}$, Ti$_4$Co$_2$O, and Ti$_4$Ir$_2$O. The value of Pauli limit $\mu_0 H_P$ is determined with a certain magnetic field at which a gain of paramagnetic Zeeman energy at a normal state is equal to a superconducting condensation energy, given in the following formula [15,16]:

$$\mu_0 H_P = \frac{\Delta(0)}{\sqrt{g}\mu_B} = 1.86 T_c, \quad (1)$$

where $g$ = 2 is a $g$-factor for free electron and $\mu_B \approx 9.27 \times 10^{-24}$ J T$^{-1}$ is a Bohr magneton. The $\Delta(0)$ is a superconducting gap energy at 0 K described as $\Delta(0) = 1.76 k_B T_c$ ($k_B \approx 1.38 \times 10^{-23}$ J K$^{-1}$ is a Boltzmann constant) in the single gap Bardeen−Cooper−Schrieffer (BCS) model [17]. The large $\mu_0 H_{c2}(0)$ overwhelming the $\mu_0 H_P$ can arise from special electronic states and structural properties such as multi-band effect [18-20], spin-triplet cooper pairing [21], Fulde-Ferrell-Larkin-Ovchinnikov (FFLO) superconducting state [22,23], global or local inversion symmetry breaking [24,25], and strong spin-orbit coupling (SOC) [26,27]. Particularly, spin-orbit scattering originating from SOC suppresses a cooper pair breaking by the Pauli paramagnetic effect because SOC destroys spin as a good quantum number, and makes spin susceptibility of the superconducting state close to that of a normal state, described in Werthamer–Helfand–Hohenberg (WHH) theory [28,29]. Therefore, the strong SOC has the potential to achieve a large $\mu_0 H_{c2}(0)$ superconducting state, and the strength can be controlled by chemical elemental substitution [30]. The strength of SOC, $\xi$ can be approximately calculated using a hydrogen-like atom model [31]:

$$\xi \propto \frac{Z^4}{n^3 l \left(l + \frac{1}{2}\right)(l+1)}, \qquad (2)$$

where $Z$, $n$, and $l$ are atomic number, principal quantum number, and orbital angular momentum, respectively. From the expression, we can understand the strength of SOC proportions to $Z^4$ within the same electronic orbital. In the case of the $Ti_4Ir_2O$ superconductor, we can expect the Ir-$5d$ orbital hosting enhanced SOC should play an important role for the large $\mu_0H_{c2}(0)$, and it was found that Ti-$3d$ and It-$5d$ orbitals hybridize near its Fermi level and the violation of the Pauli limit is a result of a combination of strong-coupled superconductivity, SOC, and strong electron correlation [32]. Furthermore, the large SOC splitting a band structure along Γ-K lines due to the Ir-$5d$ electrons was weakened by applying pressures, and large $\mu_0H_{c2}(0)$ undergoes a crossover at 35.6 GPa from well beyond to less than the $\mu_0H_P$ [33]. As mentioned above, the $\eta$-carbide-type superconductors have been studied from the points of view of the large $\mu_0H_{c2}(0)$ and SOC effect based on $d$-block transition metals. Table 1 shows a list of the $\eta$-carbide-type superconductors with $T_c$, $\mu_0H_P$, and $\mu_0H_{c2}(0)$. Some kinds of $\eta$-carbide-type oxide superconductors have been reported; however, it is not sufficient as of now, and developing new examples of them is important for a deeper understanding of the $\eta$-carbide-type oxide superconducting properties.

Herein, we focus on a $Zr_4B_2O$ system because superconductivity was solely confirmed in $Zr_4Rh_2O$ in the system to the best of our knowledge. We report the discovery of an unrevealed superconducting nature of $Zr_4Pd_2O$ known as a hydrogen storage material [34,35]. Polycrystalline samples of $Zr_4Pd_2O$ were obtained by arc melting followed by annealing in an evacuated quartz tube. We performed synchrotron X-ray diffraction (SXRD) measurement at the beamline BL13XU in SPring-8 and checked chemical composition by means of the energy dispersive X-ray spectroscopy (EDX) method to characterize obtained samples. The bulk superconductivity was confirmed through magnetic susceptibility, electrical resistivity, and specific heat measurement, resulting in $T_c$ = 2.8 K, 2.73 K, and 2.6 K, respectively. We discuss the $\mu_0H_{c2}(0)$ for $Zr_4Pd_2O$ and $Zr_4Rh_2O$ using the electrical resistivity and specific heat data measured at several magnetic fields. We find that $Zr_4Rh_2O$ shows the $\mu_0H_{c2}(0)$ = 6.16 T, lower than the $\mu_0H_P$ = 7.59 T; however, $Zr_4Pd_2O$ shows the large $\mu_0H_{c2}(0)$ = 6.88 T, violating the $\mu_0H_P$ = 5.29 T different from $Zr_4Rh_2O$. The violation of the Pauli limit for $Zr_4Pd_2O$ can be attributed to the larger strength of SOC derived from Pd-$4d$ electrons.

| Compound | $T_c$ (K) | $\mu_0H_P$ (T) | $\mu_0H_{c2}(0)$ (T) | Reference |
|---|---|---|---|---|
| $Zr_4Rh_2O_{0.7}$ | 2.8 | 5.18 | 4.89 | [12] |
| $Zr_4Rh_2O$ | 4.7<br>3.73 | 8.70<br>7.59 | 6.08<br>6.16 | [12]<br>This work |
| $Nb_4Rh_2C_{1-\delta}$ | 9.75 | 18.0 | 28.5 | [13] |
| $Ti_4Co_2O$ | 2.7 | 5.02 | 7.08 | [14] |
| $Ti_4Rh_2O$ | 2.8 | 5.21 | 5.15 | [14] |
| $Ti_4Ir_2O$ | 5.4<br>5.12<br>5 | 9.86<br>9.52<br>9.58 | 16.06<br>16.45<br>18.2 | [14]<br>[32]<br>[33] |
| $Zr_4Pd_2O$ | 2.73 | 5.29 | 6.88 | This work |

**Table 1.** A list of $\eta$-carbide-type superconductors. The superconducting transition temperature $T_c$, Pauli limit $\mu_0H_P$, and upper critical field $\mu_0H_{c2}(0)$ are shown.

## Results
### Crystal structures of Zr₄Pd₂O and Zr₄Rh₂O

Schematic images of the crystal structure for $Zr_4Tr_2O$ ($Tr$ = Pd or Rh) are shown in Fig. 1(a). These compounds crystalline a cubic $\eta$-carbide crystal structure with a space group $Fd\bar{3}m$ (No. 227). The metal atoms Zr occupy Wyckoff positions 48f (labeled as Zr1), 16d (labeled as Zr2) and $Tr$ occupies Wyckoff position 32e position. The O atom occupies Wyckoff position 16c, and the occupying can be regarded as void filling in a Ti$_2$Ni-type structure. The complicated $\eta$-carbide crystal structure consists of Zr1 octahedra centered by O (Fig.1 (b)) and a geometrically frustrated stella quadrangula lattice (Fig.1 (c)) [36]. The Zr1 octahedra caging the O atom at the center are arranged in the unit cell sharing the corner as seen in a pyrochlore structure. The stella quadrangula lattice can be formed by inserting a small tetrahedron into each tetrahedron making the pyrochlore lattice, and the unit consists of nested $Tr$ and Zr2 tetrahedra with the same center of gravity [37,38]. The SXRD patterns and Rietveld refinement results of Zr$_4$Pd$_2$O and Zr$_4$Rh$_2$O at Room temperature are shown in Figs. 1(d) and 1(e), respectively. The SXRD patterns were well fit to cubic $\eta$-carbide crystal structure and the lattice constants of Zr$_4$Pd$_2$O and Zr$_4$Rh$_2$O were determined to $a$ = 12.4617(1) Å and 12.3977(3) Å, respectively. The values of $a$ for Zr$_4$Pd$_2$O and Zr$_4$Rh$_2$O were confirmed to be close to the previous report [12,34,35]. We found a small amount of impurity phases such as ZrO$_2$ and Zr in both of them, and reliability factors $R_{wp}$ were 8.438% for Zr$_4$Pd$_2$O and 16.550% for Zr$_4$Rh$_2$O. We also refined the atomic coordinates and isotropic atomic displacement parameter $U_{iso}$ for each atom, as summarized in Table 2. The $U_{iso}$ of the O atom was fixed to be 0.004 because the value approximately close to be zero within the errors on the fitting quality. The chemical compositions of Zr$_4$Pd$_2$O and Zr$_4$Rh$_2$O confirmed through EDX were to be 2.10(2) for Zr:Pd and 2.2(2) for Zr:Rh.

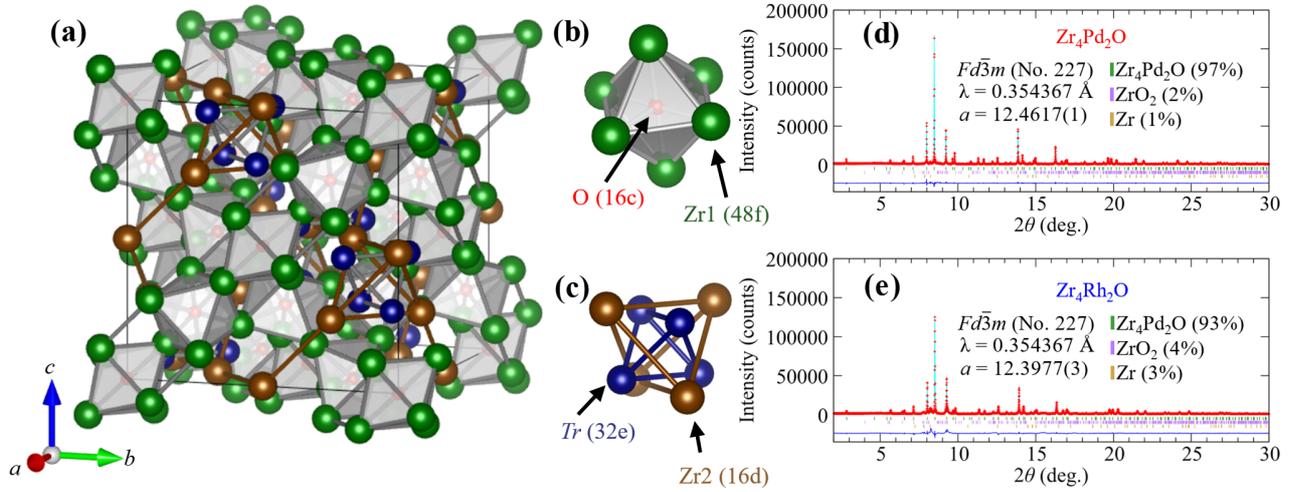

**Figure 1.** Schematic images of the cubic $\eta$-carbide crystal structure. (**a**) Zr$_4$$Tr$$_2$O ($Tr$ = Pd or Rh) unit cell. (**b**) Zr1 octahedron centered by O atom. (**c**) Unit of stella quadrangla lattice. Rietveld refined Room temperature SXRD patterns of (**d**) Zr$_4$Pd$_2$O and (**e**) Zr$_4$Rh$_2$O. The red points and cyan lines represent obtained SXRD data and calculated data, respectively. Lower solid lines show the Bragg peak positions. The lower blue lines are differences between obtained SXRD data and calculated data.

| Compound | Zr$_4$Pd$_2$O | Zr$_4$Rh$_2$O |
| --- | --- | --- |
| space group | $Fd\bar{3}m$ (No. 227) | |
| Z | 16 | |
| $a$ (Å) | 12.4617(1) | 12.3977(3) |
| $V$ (Å$^3$) | 1935.22(3) | 1905.54(9) |
| $x$ (Zr1) | 0.30762(7) | 0.3111(2) |
| $x$ ($Tr$) | 0.28569(5) | 0.2815(1) |
| $U_{iso}$ (Zr1) | 0.0051(2) | 0.0052(5) |
| $U_{iso}$ (Zr2) | 0.0104(4) | 0.017(1) |
| $U_{iso}$ ($Tr$) | 0.0126(3) | 0.0193(8) |
| $U_{iso}$ (O) | 0.004 (fixed) | 0.004 (fixed) |
| $R_{wp}$ (%) | 8.438 | 16.550 |

**Table 2.** Crystalline parameters obtained from Rietveld refinement using SXRD patterns at Room temperature for Zr$_4$Pd$_2$O and Zr$_4$Rh$_2$O. The atomic coordinates of Zr1, Zr2, $Tr$, and O are ($x$,1/8,1/8), (1/2,1/2,1/2), ($x$,$x$,$x$), and (0,0,0), respectively.

### Magnetization, Electrical resistivity, and Specific heat

We measured temperature- and magnetic field-dependent magnetizations for Zr$_4$Pd$_2$O and Zr$_4$Rh$_2$O with polished rectangular cuboid samples. The value of a demagnetizing factor $N$ with a rectangular cuboid sample applied a vertical magnetic field can be calculated using dimensional information of the sample: length $l$, width $w$, and thickness $t$ [39]:

$$N = \frac{4lw}{4lw + 3t(l + w)}. \tag{3}$$

The calculated values of $N$ were 0.66 for $Zr_4Pd_2O$ ($l$ = 1.40 mm, $w$ = 1.50 mm, $t$ = 0.51 mm) and 0.81 for $Zr_4Rh_2O$ ($l$ = 0.96 mm, $w$ = 1.37 mm, $t$ = 0.18 mm). An actual magnetic field in samples should be modified to an effective inner magnetic field as described in $H_{eff} = H - 4\pi MN$, where $H$ is an applied magnetic field and $M$ is a magnetization. Thus, magnetic susceptibility $\chi$ taken to account for the demagnetizing effect is defined as follows:

$$\chi = \frac{M}{H_{eff}} = \frac{M}{H - 4\pi MN}. \tag{4}$$

Figures 2(a) and 2(b) show temperature dependences of the $\chi$ for $Zr_4Pd_2O$ and $Zr_4Rh_2O$, respectively. We observed a clear superconducting transition at $T_c$ = 2.8 K for $Zr_4Pd_2O$ and $T_c$ = 3.5 K for $Zr_4Rh_2O$. The temperature width in the superconducting transition for $Zr_4Rh_2O$ was broader than $Zr_4Pd_2O$. Moreover, the $T_c$ of that was lower than the previous report ($T_c$ = 4.3 K) and it would be based on the vacancy of oxygen [12]. The superconducting state at 1.8 K reached perfect diamagnetism in a zero-field cooling (ZFC) process different from a case for field cooling (FC) process. The hysteresis of temperature-dependent $\chi$ with the cooling process reflects the nature of type-II superconductor. Temperature dependence of a lower critical field $\mu_0H_{c1}$ can be obtained through magnetic field dependence of $M$ as shown in Figs. 2(c) and 2(d). The horizontal axis is displayed in $\mu_0H_{eff}$ instead of $H$ for precise estimation of the lower critical field. In $Zr_4Pd_2O$, the measurement was carried out in a range of 1.8 K $< T <$ 2.8 K with increments of 0.1 K. In the case of $Zr_4Rh_2O$, the data was taken at 1.8 K, 1.9 K, and 2.0 K, then taken with increments of 0.2 K up to 3.6 K. We observed the convex downward curve of $M$ as functions of $\mu_0H_{eff}$ at each temperature, and the minimum points gradually shifted lower filed as increasing temperature. In a low-filed region, the linear responses of $M$ corresponding to the Meissner state were observed, and the behavior can be described as $M_{fit} = a^*H_{eff} + b^*$, where $a^*$ and $b^*$ are numerical constants. The fitting was carried out in a range of 0 mT $< \mu_0H_{eff} <$ 5 mT. Figures 2(e) and 2(f) are differences between $M$ and $M_{fit}$ for $Zr_4Pd_2O$ and $Zr_4Rh_2O$, respectively. The dashed lines represent the value of $\mu_0H_{eff}$ which deviates from the linear behavior of $M$. Temperature dependences of lower critical field $\mu_0H_{c1}(T)$ collected from the $M$-$M_{fit}$ are shown in Figs. 2(g) and 2(h) for $Zr_4Pd_2O$ and $Zr_4Rh_2O$, respectively. The lower critical field at 0 K, $\mu_0H_{c1}(0)$ can be obtained from the following empirical formula:

$$\mu_0H_{c1}(T) = \mu_0H_{c1}(0)\left[1 - \left(\frac{T}{T_c}\right)^2\right], \tag{5}$$

resulting $\mu_0H_{c1}(0)$ = 11.3 mT and 8.2 mT for $Zr_4Pd_2O$ and $Zr_4Rh_2O$, respectively. $Zr_4Pd_2O$ showed a higher $\mu_0H_{c1}(0)$ than that of $Zr_4Rh_2O$ even lower $T_c$, suggesting that $Zr_4Pd_2O$ tends to be more robust against magnetic field rather than $Zr_4Rh_2O$.

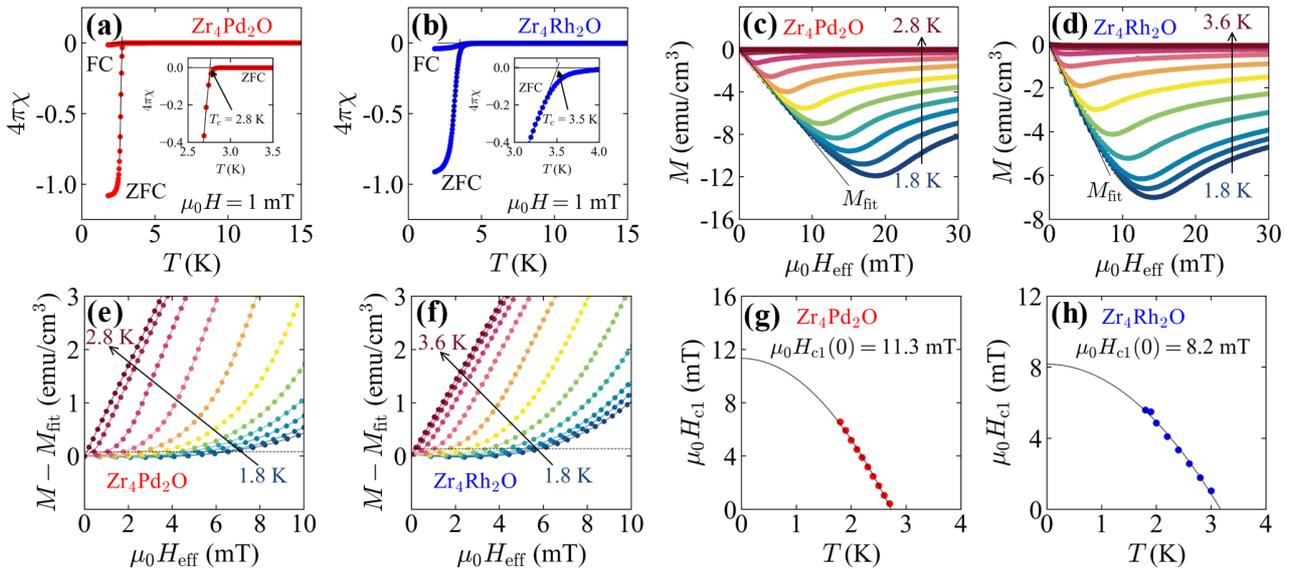

**Figure 2.** (**a**,**b**) ZFC and FC temperature-dependent magnetic susceptibility under $\mu_0H$ = 1 mT for (**a**) $Zr_4Pd_2O$ and (**b**) $Zr_4Rh_2O$. The insets are enlarged view near $T_c$. (**c**,**d**) Effective inner magnetic field dependence of magnetic susceptibility for (**c**) $Zr_4Pd_2O$ and (**d**) $Zr_4Rh_2O$. The solid lines are fit of $M_{fit} = a^*H_{eff} + b^*$ in a range of 0 mT $< \mu_0H_{eff} <$ 5 mT. (**e**,**f**) A difference between $M$ and $M_{fit}$ for (**e**) $Zr_4Pd_2O$ and (**f**) $Zr_4Rh_2O$. The dashed lines correspond to the field where $M$ begins to deviate from the linear behavior. The dashed lines are used to determine temperature dependence of the lower critical field. (**g**,**h**) Temperature

dependence of lower critical field for (**g**) $Zr_4Pd_2O$ and (**h**) $Zr_4Rh_2O$. The solid lines are the fit of $\mu_0H_{c1}(T) = \mu_0H_{c1}(0)[1-(T/T_c)^2]$. The values of $\mu_0H_{c1}(0)$ were calculated to be 11.3 mT and 8.2 mT for $Zr_4Pd_2O$ and $Zr_4Rh_2O$, respectively.

Temperature dependences of electrical resistivity $\rho(T)$ for $Zr_4Pd_2O$ and $Zr_4Rh_2O$ at zero field are shown in Figs. 3(a) and 3(b), respectively. In a low-temperature region, we observed a drop of $\rho(T)$ to zero, suggesting a superconducting transition. $Zr_4Pd_2O$ showed a sharp transition, and the zero resistivity was observed at $T_c^{zero}$ = 2.73 K. $Zr_4Rh_2O$, however, showed a broad transition, and the zero resistivity was observed at $T_c^{zero}$ = 3.73 K. The $T_c$ obtained from the zero resistivity agreed with the result from the magnetic susceptibility measurement. The $\rho(T)$ exhibits a metallic behavior in a normal state for both $Zr_4Pd_2O$ and $Zr_4Rh_2O$. In the low-temperature normal state, the $\rho(T)$ curve can be fitted using the power-law model:

$$\rho(T) = \rho_0 + AT^{n_{PL}}, \tag{6}$$

where $\rho_0$, $A$, and $n_{PL}$ are residual resistivity, temperature-independent coefficient, and power exponent respectively. The $\rho(T)$ curve of $Zr_4Pd_2O$ in a range of 4 K < $T$ < 80 K was fitted using the model, providing $\rho_0$ = 0.22 mΩ cm, $A$ = 0.00011 mΩ K$^{-2}$, and $n_{PL}$ = 1.2. For the $\rho(T)$ curve of $Zr_4Rh_2O$, we fitted in a range of 6 K < $T$ < 80 K, obtained $\rho_0$ = 0.29 mΩ cm, $A$ = 0.00017 mΩ K$^{-2}$, and $n_{PL}$ = 1.3. The values of $n_{PL}$ close to 1 suggest that the $\rho(T)$ show linear-temperature dependence in the low-temperature normal state for $Zr_4Pd_2O$ and $Zr_4Rh_2O$, and the behavior was observed in some $\eta$-carbide-type superconductors [12,13]. In a high-temperature region (80 K < $T$ < 300 K) where electron-phonon interaction is a dominant mechanism of electron scattering, the $\rho(T)$ shows a convex upward curve with increasing temperature. A similar trend can be found in many superconductors consisting of $d$-block element [40-43], and the convex upward curve of $\rho(T)$ in the high-temperature region can be fitted using the following parallel resistor model, yielded by Wiesmann et al. [44]:

$$\rho(T) = \left[\frac{1}{\rho_{sat}} + \frac{1}{\rho_{ideal}(T)}\right]^{-1}. \tag{7}$$

The temperature-independent term $\rho_{sat}$ corresponds to the saturation of $\rho(T)$ in high temperature. Fisk and Webb found the saturation of resistivity in A-15 superconductors such as $Nb_3Sn$ [45], and the saturation can be realized when a mean free path becomes comparable to interatomic separations of the material, called as Ioffe-Regel condition [46]. The temperature-dependent component $\rho_{ideal}$ is described with the Bloch-Grüneisen model [47]:

$$\rho_{ideal}(T) = \rho_{ideal,0} + B\left(\frac{T}{\Theta_D}\right)^{n_{BG}} \int_0^{\frac{\Theta_D}{T}} \frac{t^{n_{BG}}}{(e^t - 1)(1-e^{-t})} dt, \tag{8}$$

where $\rho_{ideal,0}$, $B$, $\Theta_D$, and $n_{BG}$ are ideal temperature-independent residual resistivity, temperature-independent coefficient, Debye temperature, and power exponent with the Bloch-Grüneisen model, respectively. The $n_{BG}$ usually takes 2, 3, or 5 depending on the scattering nature. The best fit was obtained when $n_{BG}$ = 5 for both $Zr_4Pd_2O$ and $Zr_4Rh_2O$, and the calculation provided $\rho_{sat}$ = 0.52 mΩ cm, $\rho_{ideal,0}$ = 0.41 mΩ cm, $B$ = 1.49 mΩ cm, and $\Theta_D$ = 261 K for $Zr_4Pd_2O$, and $\rho_{sat}$ = 0.82 mΩ cm, $\rho_{ideal,0}$ = 0.49 mΩ cm, $B$ = 1.68 mΩ cm, and $\Theta_D$ = 198 K for $Zr_4Rh_2O$. The $\rho_0$ can be calculated using $\rho_{sat}$ and $\rho_{ideal,0}$ as given in $\rho_0 = \rho_{sat}\rho_{ideal,0}/(\rho_{sat} + \rho_{ideal,0})$, and the obtained values of $\rho_0$ are 0.23 mΩ cm and 0.31 mΩ cm, for $Zr_4Pd_2O$ and $Zr_4Rh_2O$, respectively. These $\rho_0$ values are consistent with those obtained by the power-law model. The fitted curves using the power-law model and parallel resistor model are displayed as dashed lines and solid lines, respectively, in Figs. 3(a) and 3(b). Resistivity at 300 K, $\rho_{300\,K}$, was found to be 0.32 mΩ cm for $Zr_4Pd_2O$ and 0.47 mΩ cm for $Zr_4Rh_2O$. Residual resistivity ratio, RRR = $\rho_{300\,K}/\rho_0$ was calculated to be 1.42 and 1.61 for $Zr_4Pd_2O$ and $Zr_4Rh_2O$, respectively. The small RRR value is also seen in other $\eta$-carbide-type superconductors [12-14], and the poor metallic behavior is a common feature of polycrystalline metallic oxide compounds whose grain-boundary scattering is significant [48]. Figures 3(c) and 3(d) show the $\rho(T)$ curves at several magnetic fields for $Zr_4Pd_2O$ and $Zr_4Rh_2O$, respectively. The magnetic fields are applied with an increment of 0.2 T for $Zr_4Pd_2O$ up to $\mu_0H$ = 3.8 T. For $Zr_4Rh_2O$, the magnetic fields are increased by 0.2 T up to $\mu_0H$ = 4.0 T and then increased by 0.5 T up to $\mu_0H$ = 6.5 T. The $T_c^{zero}$ shifted lower temperature with increasing magnetic field as we expected. We used typical 10%, 50%, and 90% criteria defined with $\rho_0$ to determine temperature dependences of the upper critical field for $Zr_4Pd_2O$ and $Zr_4Rh_2O$ (discussed later).

Figures 3(e) and 3(f) show temperature dependences of total specific heat $C(T)$ at several magnetic fields for $Zr_4Pd_2O$ and $Zr_4Rh_2O$, respectively. Magnetic fields were applied with increments of 0.2 T up to $\mu_0H$ = 2 T and also measured at $\mu_0H$ = 9 T. We observed clear specific heat jumps, suggesting superconducting transition, up to $\mu_0H$ = 2 T, and the temperature at which the jumps observed shifted to a lower temperature, consistent with the results from electrical resistivity. $Zr_4Rh_2O$ showed broader transitions than that of $Zr_4Pd_2O$, as seen in magnetic susceptibility and electrical resistivity measurements. To calculate Sommerfeld coefficient $\gamma$ and $\Theta_D$, we fitted $C(T)$ using the following formula:

$$C(T) = \gamma T + \beta T^3 + \delta T^5, \tag{9}$$

where $\beta$ and $\delta$ are the coefficients of the phonon contributions for the harmonic and anharmonic terms, respectively. As a result of the fitting, we obtained the values of $\gamma$ to be $\gamma$ = 32.5 mJ K$^{-2}$ mol$^{-1}$ for Zr$_4$Pd$_2$O and $\gamma$ = 18.1 mJ K$^{-2}$ mol$^{-1}$ for Zr$_4$Rh$_2$O. For the phonon contribution coefficients, we obtained $\beta$ = 1.87 mJ K$^{-4}$ mol$^{-1}$ and $\delta$ = 0.0016 mJ K$^{-6}$ mol$^{-1}$ for Zr$_4$Pd$_2$O, and $\beta$ = 1.12 mJ K$^{-4}$ mol$^{-1}$ and $\delta$ = 0.014 mJ K$^{-6}$ mol$^{-1}$ for Zr$_4$Rh$_2$O. The fitting curves are shown as solid lines in Figs. 3(e) and 3(f). The values of $\Theta_D$ can be calculated using the $\beta$ as the following formula:

$$\Theta_D = \left(\frac{12\pi^4 NR}{5\beta}\right)^{\frac{1}{3}}, \quad (10)$$

where $N$ = 7 is the number of atoms per formula unit and $R \approx$ 8.31 J K$^{-1}$ mol$^{-1}$ is an ideal gas constant. The calculated $\Theta_D$ was $\Theta_D$ = 194 K and 230 K for Zr$_4$Pd$_2$O and Zr$_4$Rh$_2$O, respectively, and the values were close to the calculation result obtained by the parallel resistor model in electrical resistivity measurement. Temperature dependences of the electron contribution of the specific heat $C_{el}(T)$ estimated by subtracting phonon contributions $\beta T^3 + \delta T^5$ from $C(T)$ are shown in Figs. 3(g) and 3(h) for Zr$_4$Pd$_2$O and Zr$_4$Rh$_2$O, respectively. $T_c$ determined from $C_{el}(T)$ at zero field was 2.6 K for Zr$_4$Pd$_2$O and 3.3 K for Zr$_4$Rh$_2$O. The normalized jumps of $C_{el}(T)$, $\Delta C_{el}/\gamma T_c$, were estimated to be 1.58 and 1.57 for Zr$_4$Pd$_2$O and Zr$_4$Rh$_2$O, respectively. The values of the jump were similar and slightly higher than 1.43, which is the expected value by the weak-coupling BCS theory [17]. This result suggests that Zr$_4$Pd$_2$O and Zr$_4$Rh$_2$O are electron-phonon coupling superconductors with a little strong-coupling nature. We can calculate an electron-phonon coupling constant $\lambda_{el-ph}$ using the McMillan formula [49]:

$$\lambda_{el-ph} = \frac{1.04 + \mu^* \ln\left(\frac{\Theta_D}{1.45 T_c}\right)}{(1 - 0.62\mu^*) \ln\left(\frac{\Theta_D}{1.45 T_c}\right) - 1.04}, \quad (11)$$

where $\mu^*$ = 0.13 is a Coulomb coupling constant and the value is used empirically for similar materials containing transition metals. We obtained the values of $\lambda_{el-ph}$ to be 0.60 for Zr$_4$Pd$_2$O and 0.61 for Zr$_4$Rh$_2$O. An electronic density of states at the Fermi energy $D(E_F)$ is proportional to a term $(1+\lambda_{el-ph})$ when we consider the electron-phonon coupling. Therefore, $D(E_F)$ with spin degeneracy can be expressed in the following:

$$D(E_F) = \frac{3\gamma}{\pi^2 k_B^2 (1 + \lambda_{el-ph})}. \quad (12)$$

The measured $\gamma$ and calculated $\lambda_{el-ph}$ provide $D(E_F)$ = 8.61 states eV$^{-1}$ per formula unit (f.u.) and 4.76 states eV$^{-1}$ per f.u. for Zr$_4$Pd$_2$O and Zr$_4$Rh$_2$O, respectively.

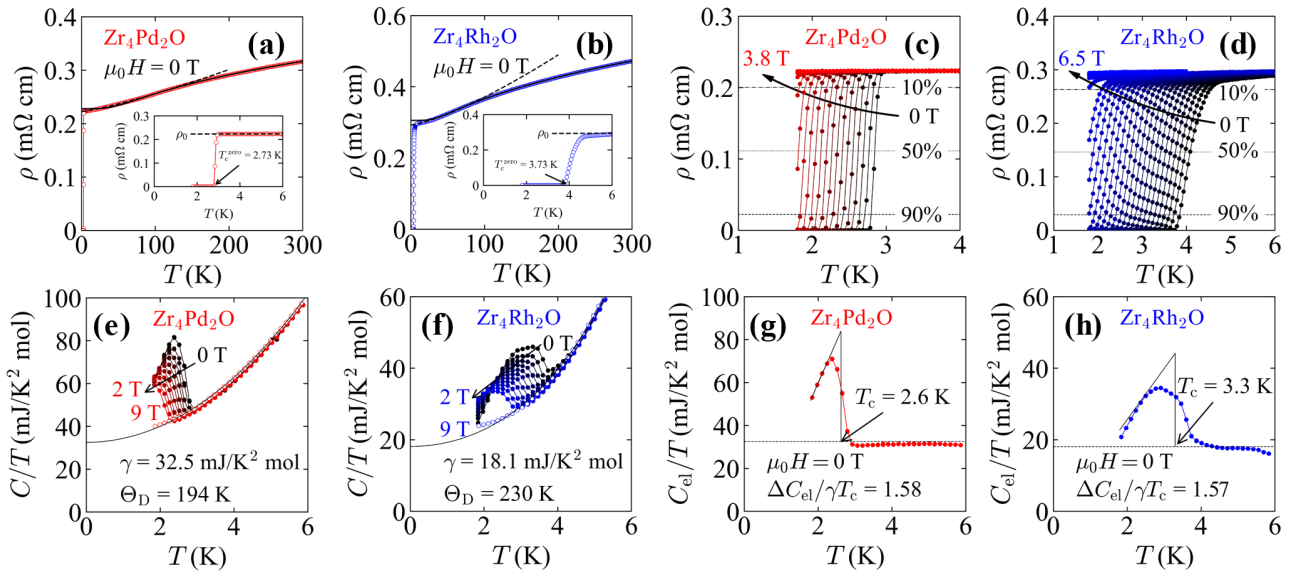

**Figure 3.** (**a,b**) Temperature dependences of electrical resistivity under zero field for (**a**) Zr$_4$Pd$_2$O and (**b**) Zr$_4$Rh$_2$O. The insets are enlarged view near $T_c$. The solid and dashed lines are fit to parallel resistor model and power-law model, respectively. (**c,d**) Temperature dependences of electrical resistivity under several magnetic fields for (**c**) Zr$_4$Pd$_2$O and (**d**) Zr$_4$Rh$_2$O. The dashed lines represent the 10%, 50%, and 90% criteria to determine temperature dependence of the upper critical field. (**e,f**) Temperature dependences of total specific heat under several magnetic fields for (**e**) Zr$_4$Pd$_2$O and (**f**) Zr$_4$Rh$_2$O. The solid lines

are fit to $C(T)/T = \gamma+\beta T^2+\delta T^4$. (**g,h**) Temperature dependences of electronic specific heat zero field for (**g**) $Zr_4Pd_2O$ and (**h**) $Zr_4Rh_2O$. The solid lines are used to estimate $T_c$ and dashed lines represent $\gamma$ value.

## Discussion

Here, we discuss the upper critical fields and other superconducting parameters of $Zr_4Pd_2O$ and $Zr_4Rh_2O$. Figures 4(a) and 4(b) are temperature dependences of upper critical field $\mu_0H_{c2}(T)$ for $Zr_4Pd_2O$ and $Zr_4Rh_2O$, respectively. The data points were taken from temperature dependences of $\rho(T)$ with 10%, 50%, and 90% criteria, and $C(T)$ under several magnetic fields. The upper critical field at 0 K, $\mu_0H_{c2}(0)$ can be calculated by fitting the data using the Ginzburg-Landau (GL) model:

$$\mu_0H_{c2}(T) = \mu_0H_{c2}(0)\left[\frac{1-\left(\frac{T}{T_c}\right)^2}{1+\left(\frac{T}{T_c}\right)^2}\right]. \tag{13}$$

We obtained the values of $\mu_0H_{c2}(0)$ for $Zr_4Pd_2O$ to be 7.18 T for $\rho(T)$ 10% criterion, 6.88 T for $\rho(T)$ 50% criterion, 6.72 T for $\rho(T)$ 90% criterion, and 9.17 T for $C(T)$. For $Zr_4Rh_2O$, the obtained $\mu_0H_{c2}(0)$ to be 6.27 T for $\rho(T)$ 10% criterion, 6.16 T for $\rho(T)$ 50% criterion, 5.91 T for $\rho(T)$ 90% criterion, and 7.74 T for $C(T)$. We found that the whole values of $\mu_0H_{c2}(0)$ for $Zr_4Pd_2O$ were higher than that of $\mu_0H_P = 5.29$ T calculated with $T_c$ of $\rho(T)$ 50% criterion. On the other hand, for $Zr_4Rh_2O$, the values of $\mu_0H_{c2}(0)$ derived from $\rho(T)$ criterion were lower than that of $\mu_0H_P = 7.59$ T calculated with $T_c$ of $\rho(T)$ 50% criterion. The value of $\mu_0H_{c2}(0)$ derived from $C(T)$ was close to the $\mu_0H_P$. The absence of violation of the Pauli limit for $Zr_4Rh_2O$ is consistent with the previous study [12]. The violation of the Pauli limit observed in $Zr_4Pd_2O$ is an unreported superconducting nature, and a similar violation was reported in other $\eta$-carbide-type superconductors as mentioned in the Introduction part. A quasiparticle mean free path $l$ at a normal state near the superconducting state can be estimated using the following formula derived from Singh et al. [50]:

$$l = 2.372\times10^{-14} \frac{\left(\frac{m^*}{m_e}\right)^2 V_M^2}{D(E_F)^2\rho_0}, \tag{14}$$

where $m_e$, $m^*$, and $V_M$ are free-electron mass, effective mass of the individual quasiparticles, and molar volume. The $l$ is in cm unit when we take $V_M$, $D(E_F)$, and $\rho_0$ are in cm$^3$ mol$^{-1}$, states eV$^{-1}$ per f.u., and $\Omega$ cm, respectively. If we assume $m^*/m = 1$, we obtain $l = 0.76$ Å for $Zr_4Pd_2O$ using $V_M = 72.8$ cm$^3$ mol$^{-1}$, $D(E_F) = 8.61$ states eV$^{-1}$ per f.u., and $\rho_0 = 0.22$ m$\Omega$ cm. Likewise for $Zr_4Rh_2O$, we obtain $l = 1.84$ Å using $V_M = 71.7$ cm$^3$ mol$^{-1}$, $D(E_F) = 4.76$ states eV$^{-1}$ per f.u., and $\rho_0 = 0.29$ m$\Omega$ cm. A GL coherence length $\xi_{GL}$ can be calculated using the GL model with $\mu_0H_{c2}(0)$ as the following:

$$\mu_0H_{c2}(0) = \frac{\Phi_0}{2\pi\xi_{GL}^2}, \tag{15}$$

where $\Phi_0 \approx 2.07 \times 10^{-15}$ Wb is a magnetic flux quantum. The values of $\xi_{GL}$ for $Zr_4Pd_2O$ and $Zr_4Rh_2O$ were calculated to be 69 Å and 73 Å, respectively, using the $\mu_0H_{c2}(0)$ obtained from $\rho(T)$ 50% criterion. The values of $\xi_{GL}$ were found to be much longer than that of $l$ for both $Zr_4Pd_2O$ and $Zr_4Rh_2O$. Therefore, both $Zr_4Pd_2O$ and $Zr_4Rh_2O$ are supposed to be in the dirty limit. The orbital limit $\mu_0H_{orb}$ can be estimated with WHH theory without considering spin-orbit scattering [28,29]. In the dirty limit, $\mu_0H_{orb}$ is expressed in the following formula:

$$\mu_0H_{orb} = -0.693T_c\left.\frac{d\mu_0H_{c2}(T)}{dT}\right|_{T=T_c}. \tag{16}$$

The slope of $\mu_0H_{c2}(T)$ at $T_c$ was estimated to be -2.72 TK$^{-1}$ and -1.81 TK$^{-1}$ for $Zr_4Pd_2O$ and $Zr_4Rh_2O$, respectively, when using the $\mu_0H_{c2}(T)$ data of the $\rho(T)$ 50% criterion. These obtained values yielded $\mu_0H_{orb} = 5.36$ T for $Zr_4Pd_2O$ and 5.11 T for $Zr_4Rh_2O$. For $Zr_4Pd_2O$, the value of $\mu_0H_{c2}(0)$ determined by $\rho(T)$ 50% criterion was found to be larger than that of both $\mu_0H_p$ and $\mu_0H_{orb}$. On the other hand, for $Zr_4Rh_2O$, the value of $\mu_0H_{c2}(0)$ determined by $\rho(T)$ 50% criterion was higher than that of $\mu_0H_{orb}$ but lower than that of $\mu_0H_p$. The enhanced $\mu_0H_{c2}(0)$ larger than both $\mu_0H_p$ and $\mu_0H_{orb}$ for $Zr_4Pd_2O$ implies the importance of spin-orbit scattering caused by strong SOC because the strong SOC can suppress the Pauli paramagnetic pair-breaking effect [28,29] and calculation of $\mu_0H_{orb}$ in Eq. (16) does not consider the SOC. The importance of SOC was pointed out by Ruan et al. [32] and Shi et al. [33] in $Ti_4Ir_2O$, exhibiting the violation of the Pauli limit. Similarly, we can expect that the enhanced $\mu_0H_{c2}(0)$ of $Zr_4Pd_2O$ would be raised from strong SOC. The strength of SOC proportions to $Z^4$ within the same electronic orbital as expressed in Eq. (2), therefore the absence of enhanced $\mu_0H_{c2}(0)$ for $Zr_4Rh_2O$ may be explained by the lower strength of SOC because of the number of $d$ electron configuration: Rh consists of $4d^8$, but Pd consists of $4d^{10}$. A GL penetration depth $\lambda_{GL}$ can be obtained using $\mu_0H_{c1}$ and $\xi_{GL}$ in the following formula:

$$\mu_0 H_{c1}(0) = \frac{\Phi_0}{4\pi\lambda_{GL}^2} \ln\left(\frac{\lambda_{GL}}{\xi_{GL}}\right). \tag{17}$$

We obtained the values of $\lambda_{GL}$ to be 2250 Å for $Zr_4Pd_2O$ and 2697 Å for $Zr_4Rh_2O$. GL parameters $\kappa_{GL} = \lambda_{GL}/\xi_{GL}$ were estimated to be 33 and 37 for $Zr_4Pd_2O$ and $Zr_4Rh_2O$, respectively. The calculation results agree with the nature of the type-II superconductor, shown in Figs. 2(a) and 2(b). A thermodynamic critical field $\mu_0 H_c$ can be estimated using the following expression:

$$\mu_0 H_{c1}(0) \cdot \mu_0 H_{c2}(0) = [\mu_0 H_c(0)]^2 \ln\kappa_{GL}. \tag{18}$$

The calculation provided the values of $\mu_0 H_c(0)$ to be 150 mT and 118 mT for $Zr_4Pd_2O$ and $Zr_4Rh_2O$, respectively. Finally, we summarized the whole obtained superconducting properties in Table 3.

In summary, we have discovered the bulk superconductivity in $Zr_4Pd_2O$. The crystal structure was found to be the $\eta$-carbide-type structure with a space group $Fd\bar{3}m$ (No. 227) through SXRD measurement. The bulk superconductivity was measured by magnetic susceptibility, electrical resistivity, and specific heat measurement, resulting in $T_c$ = 2.8 K, 2.73 K, and 2.6 K, respectively. $Zr_4Pd_2O$ was found to belong to the type-II superconductor by magnetic susceptibility measurement in ZFC and FC processes. The upper critical field was determined from electrical resistivity and specific heat data under several magnetic fields. We found that $Zr_4Pd_2O$ exhibited an enhanced upper critical field $\mu_0 H_{c2}(0)$ = 6.72 T violating the Pauli limit $\mu_0 H_P$ = 5.29 T, whereas the absence of the property in isostructural $\eta$-carbide-type oxide superconductor $Zr_4Rh_2O$. The enhanced upper critical field can be raised from strong SOC.

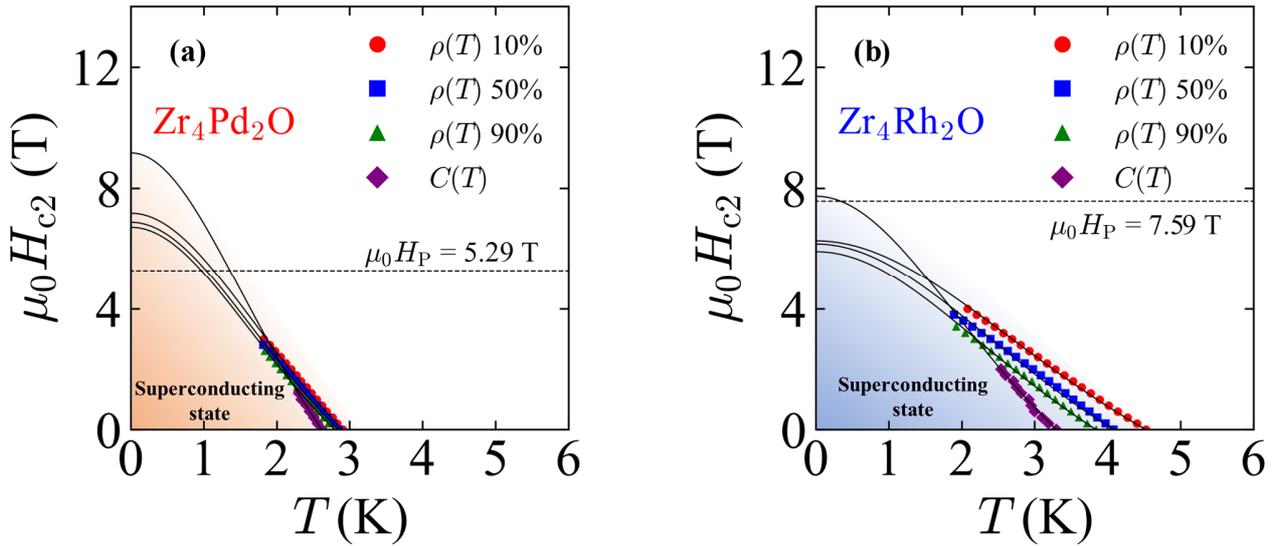

**Figure 4.** (**a**,**b**) Temperature dependence of upper critical field for (**a**) $Zr_4Pd_2O$ and (**b**) $Zr_4Rh_2O$. The solid lines are fit to the GL model. The value of $\mu_0 H_P$ was calculated using $\mu_0 H_P = 1.86T_c$ with $\rho(T)$ 50% criteria data.

| Parameters | Units | $Zr_4Pd_2O$ | $Zr_4Rh_2O$ |
|---|---|---|---|
| $T_c^{\text{magnetization}}$ | K | 2.8 | 3.5 |
| $T_c^{\text{zero resistivity}}$ | K | 2.73 | 3.73 |
| $T_c^{\text{specific heat}}$ | K | 2.6 | 3.3 |
| $\mu_0 H_{c1}(0)$ | mT | 11.3 | 8.2 |
| $\mu_0 H_{c2}(0)$ | T | 6.72 | 6.16 |
| $\mu_0 H_c(0)$ | mT | 150 | 118 |
| $\mu_0 H_P$ | T | 5.29 | 7.59 |
| $\mu_0 H_{\text{orb}}$ | T | 5.36 | 5.11 |
| $\xi_{GL}$ | Å | 69 | 73 |
| $\lambda_{GL}$ | Å | 2250 | 2697 |
| $\kappa_{GL}$ | - | 33 | 37 |
| $\rho_0$ | mΩ cm | 0.22 | 0.29 |
| $\Theta_D$ | K | 194 | 230 |
| $\gamma$ | mJ K$^{-2}$ mol$^{-1}$ | 32.5 | 18.1 |
| $\Delta C_{el}/\gamma T_c$ | - | 1.58 | 1.57 |
| $\lambda_{\text{el-ph}}$ | - | 0.60 | 0.61 |
| $D(E_F)$ | states eV$^{-1}$ per f.u. | 8.61 | 4.76 |
| $l$ | Å | 0.76 | 1.84 |

**Table 3.** The measured superconducting properties of $Zr_4Pd_2O$ and $Zr_4Rh_2O$.

## Methods

### Sample preparation

Polycrystalline samples of $Zr_4Pd_2O$ and $Zr_4Rh_2O$ were prepared by reaction of the Zr plate (99.2%, Nilaco Corporation), $ZrO_2$ powder (98.0%, Wako Special Grade), Rh powder (99.9%, Kojundo Chemical), and Pd powder (99.9%, Kojundo Chemical). These starting materials were weighed to a stoichiometric ratio, and the powders of that were pressed into a pellet. At first, the obtained pellet and Zr plate were melted together by means of an arc melting method on a water-cooled copper stage. Gas inside the arc furnace was replaced by pure argon gas 3 times and then filled with pure argon gas. Before melting the sample, a titanium ingot was melted to reduce residual oxygen gas in the furnace. The sample was melted at least 6 times and turned over at each melting for homogeneity. We observed a negligible 1–2% mass loss after the melting. Second, we crushed the as-cast sample into fine powder and pressed it into a pellet. Subsequently, we sealed the pellet into an evacuated quartz tube and treated an annealing process for 10 days at 800 °C. A mass loss was not observed after the annealing, implying oxygen in the sample was maintained.

### Crystal structure and composition

The phase purity and crystal structure of $Zr_4Pd_2O$ and $Zr_4Rh_2O$ were checked by XRD with Cu-K$\alpha$ radiation using $\theta$-$2\theta$ method. The XRD measurement was performed on a Miniflex 600 (Rigaku) equipped with a high-resolution semiconductor detector D/tex-Ultra. For further investigation, we also performed SXRD measurement at the beamline BL13XU in SPring-8 (proposal no. 2023B1669) with a wavelength of $\lambda$ = 0.354367 Å. The obtained SXRD patterns were refined by means of the Rietveld method using RIETAN-FP [51]. Schematic images of the crystal structure were depicted using VESTA [52]. The chemical compositions of Zr and *Tr* (Pd or Rh) were examined by EDX on a scanning electron microscope TM-3030plus (Hitachi High-Tech) equipped with computer software SwiftED (Oxford). The chemical composition of oxygen was not considered because of the difficulty of detecting light elements with X-ray spectroscopy.

**Measurement of superconducting properties**

Temperature and magnetic field dependence of magnetization were measured using a superconducting quantum interference device (SQUID) on a Magnetic Property Measurement System 3 (MPMS3, Quantum Design) equipped with a 7 T superconducting magnet. The measurement was performed using a vibrating sample magnetometry (VSM) mode with polished rectangular cuboid samples to estimate precise demagnetizing factors. The samples were placed in a vertically applied magnetic field. Temperature dependence was measured under $\mu_0H = 1$ mT in both zero-field cooling (ZFC) and field cooling (FC) processes. The magnetic field dependence was measured up to $\mu_0H = 30$ mT at several temperatures. Temperature and magnetic field dependence of Electrical resistivity and specific heat measurements were performed using a physical property measurement system (PPMS Dynacool, Quantum Design) equipped with a 9 T superconducting magnet. Electrical resistivity was measured by a four-probe DC method using silver paste and gold wires for the contact between a polished rectangular cuboid sample and sample puck. The measurement was performed using an excitation current of 1 mA. The specific heat measurement was carried out by means of a thermal relaxation method. The sample was mounted on a stage with N-grease for good thermal connection.

## Acknowledgements

The authors thank O. Miura for his support in experiments. This work was partly supported by JSPS-KAKENHI (No. 23KK0088), Tokyo Government Advanced Research (H31-1), TMU Research Fund for Emergent Future Society, JST-ERATO (JPMJER2201), and JST PRESTO (Grant Numbers JPMJPR21Q8).

## Author contributions statement

Y.W. and Y.M. designed the research. Y.W., A.Y., A.M., and C.M. conducted the experiment(s). Y.W. and Y.M. analyzed the results. All authors reviewed the manuscript.

## Additional information

The authors declare no competing interests.